\documentclass[prb,preprint,eqsecnum,showpacs,showkeys]{revtex4}%
\usepackage{amsmath}
\usepackage{graphicx}
\usepackage{amssymb}
\usepackage{amsfonts}%
\begin{document}
\title{Evaluation of Coulomb potential in a triclinic cell with periodic boundary conditions}
\author{Sandeep Tyagi}
\affiliation{Frankfurt Institute for Advanced Studies, J. W. Goethe Universit\"{a}t,
D-60438 Frankfurt am Main, Germany}

\begin{abstract}
Lekner [J. Lekner, Mol. Simul. \textbf{20} (1998) ] and Sperb's [R. Sperb,
Mol. Simul. \textbf{13}, (1994)] work on the evaluation of Coulomb energy and
forces under periodic boundary conditions is generalized that makes it
possible to use a triclinic unit cell in simulations in 3D rather than just an
orthorhombic cell. The expressions obtained are in a similar form as
previously obtained by Lekner and Sperb for the especial case of orthorhombic
cell. 

\end{abstract}

\maketitle

\section{Introduction}

Molecular dynamics and Monte Carlo simulations of electrically charged point
particles are indispensable in condensed matter physics. Generally, in
simulations, periodic boundary conditions (pbc) are imposed to avoid unwanted
effects of boundaries\cite{tildesley}. In presence of pbc, it becomes
necessary to include the effect of image charges while calculating interaction
energy and forces. The treatment of image charges is rather trivial for short
range forces; one simply subdivides the simulation cell into several smaller
cells, such that each of these cells is bigger than the range of interaction.
The simulation time obviously scales as $N$ for such short range potentials.
However, if the interaction is long range, such as the Coulomb interaction, it
becomes impossible to take into account all of image charges without taking
recourse to some analytical technique. One of the techniques applied
extensively in treating long range Coulomb interaction is the Ewald
method\cite{ewald}. Usually, when working with several thousand charges, the
$N^{2}$ cost of computing energy of a system carrying $N$ charges overwhelms
even supercomputers. To deal with such cases, one usually works with a variant
of the Ewald method known as PPPM\cite{pppm}. Another popular method is Fast
Multipole method\cite{fmm}. Recently another method known as the MMM\cite{mmm}
method was shown to be faster and more accurate than the PPPM\ when a high
accuracy is required, but has yet to attract the attention of researchers. Our
aim in this paper is not to consider achieving linear scaling but rather to
give a genuine alternative to the Ewald method for smaller systems.

There are two main alternatives to the Ewald method.  The first one is the so
called Lekner method\cite{lekner}, and the second one is due to
Sperb\cite{mmm, sperb1}. These methods are limited in their application in
that they were derived only for an orthorhombic simulation cell. In this
regard, these methods are not as versatile as the Ewald method that can be
applied even for a triclinic cell. However, only recently, a method was
proposed\cite{tyagijcp} that extends Sperb's method and makes it possible to
employ it even for a triclinic cell. In this paper, our aim is to directly
generalize Lekner's work on orthorhombic simulation cells to a triclinic cell.
Simple expressions will be derived to this end. However, our way of
approaching this problem will be different from that of Lekner. The end
results will of course contain Lekner's work as a special case.

\section{Generalization of Lekner method}

We consider $N$ charges $q_{i}$ contained in a triclinic simulation cell in
3-dimensional space. The index $i$ runs over $i=1$ to $N.$ We assume that the
system is periodic in all dimensions. These charges interact via the Coulomb
potential. The electrostatic energy of $N$ charges can be expressed as
\begin{equation}
E_{\text{total}}=\frac{1}{2}\sum_{i,j;i\neq j}q_{i}q_{j}G(\boldsymbol{r}%
_{i}-\boldsymbol{r}_{j})+\frac{1}{2}\sum_{i}q_{i}^{2}G_{\text{self}}%
+\frac{2\pi}{3}\left(  \sum_{i}q_{i}\boldsymbol{r}_{i}\right)  ^{2},
\label{tot}%
\end{equation}
where the position of charges in the simulation cell is denoted by
$\boldsymbol{r}_{i}$. We will obtain expressions for $G(\boldsymbol{r})$ and
$G_{\text{self}}$ in 3D in this section.

The interaction between a pair of charges, a separation $\boldsymbol{r}$
apart, goes as $\left\vert \boldsymbol{r}\right\vert ^{-1}$. Such pair wise
interactions when added under pbc lead to a diverging series, if the
simulation cell is not overall charge neutral. However, if one has a charge
neutral system, then the sum leads to a conditionally convergent series. To
give a well defined meaning to the series, one has to specify how the terms in
series are to be grouped together. Usually, one assumes that the particles
interact with a screened potential that goes as $\exp\left(  -\beta\left\vert
\boldsymbol{r}\right\vert \right)  /\left\vert \boldsymbol{r}\right\vert $;
finally the limit $\beta\rightarrow0$ is taken. This is equivalent to
introducing artificial background charges, and also to taking sums over
expanding cubes. However, this technique only leads to the intrinsic part of
the energy. A dipole term has to be added if one wants the energy in the limit
of expanding spherical shells\cite{correction}. The last term in Eq.
(\ref{tot}) represents this dipole term.

We introduce a slightly different way of including the $\beta$ factor. Instead
of working with the exponential functions, we will work with the modified
Bessel function of the second kind. Of course there is not much difference
between these two different ways as for the large arguments, the modified
Bessel functions of the second kind decay exponentially as well. We start with
the fact that the limit of $K_{1/2}\left(  \beta r\right)  $ as $\beta$ tends
to zero is given by
\begin{equation}
\lim_{\beta\rightarrow0}K_{1/2}\left(  \beta r\right)  \sim\sqrt{\frac{\pi}%
{2}}\frac{1}{\left(  \beta r\right)  ^{1/2}},
\end{equation}
that makes it possible to write%
\begin{equation}
\frac{1}{r}=\sqrt{\frac{2}{\pi}}\lim_{\beta\rightarrow0}\beta^{1/2}%
\frac{K_{1/2}\left(  \beta r\right)  }{r^{1/2}}.
\end{equation}
In a triclinic cell, the position of a charge can be specified by $x_{1},$
$x_{2}$ and $x_{3}$, where $0\leq x_{i}<l_{i}$ for $i=1,2,3.$ Here $l_{i}$
denote the lengths of the sides of the triclinic basic cell. To obtain the
interaction between a pair of charges one may assume that one of the charges
is located at the origin and the other one at $\left(  x_{1},x_{2}%
,x_{3}\right)  $. Due to the pbc, one has to also consider the interaction of
the second charge with all of the periodic images of the charge at the origin.
These periodic images are located at $\left(  l_{1}m,l_{2}n,l_{3}p\right)  ,$
where $m,n$ and $p$ are integers ranging over $-\infty$ to $+\infty.$ The
distance between the first charge and a periodic image at $\left(
l_{1}m,l_{2}n,l_{3}p\right)  $ is given by
\begin{align}
r_{m,n,p}^{2} &  =(x_{1}+l_{1}m)^{2}+\left(  x_{2}+l_{2}n\right)  ^{2}+\left(
x_{3}+l_{3}p\right)  ^{2}\nonumber\\
&  +2(x_{1}+l_{1}m)\left(  x_{2}+l_{2}n\right)  \cos\alpha\nonumber\\
&  +2\left(  x_{2}+l_{2}n\right)  \left(  x_{3}+l_{3}p\right)  \cos
\beta\nonumber\\
&  +2\left(  x_{3}+l_{3}p\right)  (x_{1}+l_{1}m)\cos\gamma.
\end{align}
The function $G$ can be expressed now as%
\begin{align}
G\left(  \boldsymbol{r}\right)   &  =\lim_{\beta\rightarrow0}G\left(
\boldsymbol{r};\beta\right)  \nonumber\\
&  =\lim_{\beta\rightarrow0}\sqrt{\frac{2}{\pi}}\beta^{1/2}\sum_{m,n,p}%
\frac{K_{1/2}\left(  \beta r_{m,n,p}\right)  }{r_{m,n,p}^{1/2}}.\label{g2}%
\end{align}
Now, we switch over to the cylindrical coordinates by defining
\begin{equation}
r_{m,n,p}^{2}=\left(  x_{np}+l_{1}m\right)  ^{2}+\rho_{n,p}^{2},
\end{equation}
where
\begin{equation}
x_{n,p}=x_{1}+\left(  x_{2}+l_{2}n\right)  \cos\alpha+\left(  x_{3}%
+l_{3}p\right)  \cos\gamma,\label{defx}%
\end{equation}
and%
\begin{align}
\rho_{n,p}^{2} &  =\left(  x_{2}+l_{2}n\right)  ^{2}\sin^{2}\alpha+\left(
x_{3}+l_{3}p\right)  ^{2}\sin^{2}\gamma\nonumber\\
&  +2\left(  x_{2}+l_{2}n\right)  \left(  x_{3}+l_{3}p\right)  \left(
\cos\beta-\cos\alpha\cos\gamma\right)  .\label{defrho}%
\end{align}
For later convenience we also define%
\[
x_{n,p}^{0}=l_{2}n\cos\alpha+l_{3}p\cos\gamma,
\]%
\begin{equation}
\rho_{n,p}^{0}=\sqrt{\left(  l_{2}n\right)  ^{2}\sin^{2}\alpha+\left(
l_{3}p\right)  ^{2}\sin^{2}\gamma+2\left(  l_{2}n\right)  \left(
l_{3}p\right)  \left(  \cos\beta-\cos\alpha\cos\gamma\right)  }%
\end{equation}
and%
\begin{align}
\rho_{0,0} &  =\left[  x_{2}^{2}\sin^{2}\alpha+x_{3}^{2}\sin^{2}\gamma
+2x_{2}x_{3}\left(  \cos\beta-\cos\alpha\cos\gamma\right)  \right]
^{1/2}\nonumber\\
x_{0,0} &  =x_{1}+x_{2}\cos\alpha+x_{3}\cos\gamma.
\end{align}
Now, as shown in the appendix, $G$ can be written as%
\begin{align}
G\left(  \boldsymbol{r}\right)   &  =\frac{2}{l_{1}}\lim_{\beta\rightarrow
0}\sum_{m}\sum_{n,p}\exp\left(  i2\pi m\frac{x_{n,p}}{l_{1}}\right)
K_{0}\left(  \sqrt{m^{2}+\beta^{2}}\rho_{n,p}\right)  \nonumber\\
&  =U\left(  \boldsymbol{r}\right)  +Q\left(  \boldsymbol{r}\right)
,\label{sb}%
\end{align}
where
\begin{equation}
U\left(  \boldsymbol{r}\right)  =\frac{2}{l_{1}}\left(  \lim_{\beta
\rightarrow0}\sum_{n,p=-\infty}^{\infty}K_{0}\left(  \beta\rho_{n,p}\right)
\right)  ,
\end{equation}
and%
\begin{equation}
Q\left(  \boldsymbol{r}\right)  =\frac{4}{l_{1}}\sum_{n,p}\sum_{m=1}^{\infty
}\,K_{0}\left(  2\pi m\frac{\rho_{n,p}}{l_{1}}\right)  \cos\left(  2\pi
m\frac{x_{n,p}}{l_{1}}\right)  .\label{q}%
\end{equation}
We note that $Q$ has excellent convergence for $\varepsilon=\rho_{0,0}%
/l_{1}>0.1.$ However, for smaller $\varepsilon$ we will modify $Q$. But before
doing that we would like to obtain an expression for $U$.  This can be done
following Sperb. For this we first express $\rho_{n,p}$ as follows%
\begin{align}
\rho_{n,p}^{2} &  =\sin^{2}\alpha\left[  x_{2}+l_{2}n+\left(  x_{3}%
+l_{3}p\right)  \left(  \frac{\cos\beta-\cos\alpha\cos\gamma}{\sin^{2}\alpha
}\right)  \right]  ^{2}\nonumber\\
&  +\left(  x_{3}+l_{3}p\right)  ^{2}\left[  \sin^{2}\gamma-\left(  \frac
{\cos\beta-\cos\alpha\cos\gamma}{\sin\alpha}\right)  ^{2}\right]  \nonumber\\
&  =\sin^{2}\alpha\left[  \left(  y_{p}+l_{2}n\right)  ^{2}+\left(
x_{3}+l_{3}p\right)  ^{2}\Omega^{2}\right]  ,\label{rhonp}%
\end{align}
where%
\begin{equation}
y_{p}=x_{2}+\left(  x_{3}+l_{3}p\right)  \zeta,\label{defy}%
\end{equation}%
\begin{equation}
\zeta=\left(  \frac{\cos\beta-\cos\alpha\cos\gamma}{\sin^{2}\alpha}\right)  ,
\end{equation}
and%
\begin{equation}
\Omega=\frac{\sqrt{1-\cos^{2}\alpha-\cos^{2}\beta-\cos^{2}\gamma+2\cos
\alpha\cos\beta\cos\gamma}}{\sin^{2}\alpha}.\label{defomega}%
\end{equation}
We note that $\zeta$ and $\Omega$ are purely geometrical factors. Also, note
the relation%
\begin{equation}
\Omega^{2}+\zeta^{2}=\frac{\sin^{2}\gamma}{\sin^{2}\alpha},
\end{equation}
which will be useful later. In order to further recast the expression for $U$
we use the identity%
\begin{align}
\sum_{n=-\infty}^{\infty}K_{0}\left[  \beta\sqrt{\left(  y+nl\right)
^{2}+x^{2}}\right]   &  =\frac{\pi}{l}\frac{\exp\left(  -\beta\left\vert
x\right\vert \right)  }{\beta}\nonumber\\
&  +\frac{2\pi}{l}\sum_{n=1}^{\infty}\frac{\exp\left(  -\left\vert
x\right\vert \sqrt{\beta^{2}+\left(  \frac{2\pi}{l}n\right)  ^{2}}\right)
}{\sqrt{\beta^{2}+\left(  \frac{2\pi}{l}n\right)  ^{2}}}\cos\left(  2\pi
n\frac{y}{l}\right)  ,\label{kexp}%
\end{align}
which implies%
\begin{equation}
\lim_{\beta\rightarrow0}\sum_{n=-\infty}^{\infty}K_{0}\left[  \beta
\sqrt{\left(  y+nl\right)  ^{2}+x^{2}}\right]  =\lim_{\beta\rightarrow0}%
\frac{\pi}{l}\frac{\exp\left(  -\beta\left\vert x\right\vert \right)  }{\beta
}-\frac{1}{2}L\left[  \frac{x}{l},\frac{y}{l}\right]  ,
\end{equation}
where we have defined%
\begin{equation}
L[x,y]=\ln\left[  1-2\exp\left(  -2\pi\left\vert x\right\vert \right)
\cos\left(  2\pi y\right)  +\exp\left(  -4\pi\left\vert x\right\vert \right)
\right]  .\label{defL}%
\end{equation}
Thus, we obtain from Eq. (\ref{q}) and (\ref{kexp})%
\begin{equation}
U\left(  \boldsymbol{r}\right)  =\frac{2\pi}{l_{1}l_{2}}\lim_{\beta
\rightarrow0}M\left(  x_{3},\beta\right)  -\frac{1}{l_{1}}\sum_{p=-\infty
}^{\infty}L\left[  \frac{x_{3}+l_{3}p}{l_{2}}\Omega,\frac{x_{p}}{l_{2}%
}\right]  .\label{b1}%
\end{equation}
The $M$ in Eq. (\ref{b1}) stands for
\begin{align}
M\left(  x_{3},\beta\right)   &  =\sum_{p=-\infty}^{\infty}\frac{\exp\left[
-\beta\Omega\sin\alpha\left\vert x_{3}+l_{3}p\right\vert \right]  }{\beta
}\nonumber\\
&  =\frac{\Omega l_{3}\sin\alpha}{2}\frac{\cosh\left[  \xi\left(
1-2\frac{\left\vert x_{3}\right\vert }{l_{3}}\right)  \right]  }{\xi\sinh\xi
},\label{b2}%
\end{align}
where $\xi=\left(  \beta\Omega l_{3}/2\right)  \sin\alpha$ and a simple
geometric sum has been carried out. The limit $\beta\rightarrow0$ can now be
carried out%
\begin{align}
\lim_{\beta\rightarrow0}M\left(  x_{3},\beta\right)  ~ &  \approx\frac{\Omega
l_{3}\sin\alpha}{2}\left[  \frac{1}{3}-2\frac{\left\vert x_{3}\right\vert
}{l_{3}}+2\left(  \frac{x_{3}}{l_{3}}\right)  ^{2}+\frac{1}{\chi^{2}}\right]
\nonumber\\
&  =\frac{\Omega l_{3}\sin\alpha}{6}\left[  1-6\frac{\left\vert x_{3}%
\right\vert }{l_{3}}+\left(  \frac{x_{3}}{l_{3}}\right)  ^{2}\right]
+\frac{2}{\beta^{2}\Omega l_{3}\sin\alpha}.\label{b3}%
\end{align}
Using Eq. (\ref{sb}), (\ref{b1}), (\ref{b2})\ and (\ref{b3}) we obtain the
following expression for the energy%
\begin{align}
G\left(  \boldsymbol{r}\right)   &  =\frac{4}{l_{1}}\sum_{n,p}\sum
_{m=1}^{\infty}\,K_{0}\left(  2\pi m\frac{\rho_{n,p}}{l_{1}}\right)
\cos\left(  2\pi m\frac{x_{n,p}}{l_{1}}\right)  -\frac{1}{l_{1}}%
\sum_{p=-\infty}^{\infty}L\left[  \frac{x_{3}+l_{3}p}{l_{2}}\Omega,\frac
{y_{p}}{l_{2}}\right]  \nonumber\\
&  +\frac{\Omega l_{3}\sin\alpha}{l_{1}l_{2}}\frac{\pi}{3}\left[
1-6\frac{\left\vert x_{3}\right\vert }{l_{3}}+\left(  \frac{x_{3}}{l_{3}%
}\right)  ^{2}\right]  ,\label{main}%
\end{align}
where we have dropped the constant factor $4\pi/(V\beta^{2})$, and
$V=l_{1}l_{2}l_{3}$ $\Omega\sin\alpha$ denotes the volume of the basic
simulation cell. This dropping of the constant factor is justified on account
of charge neutrality: The overall contribution to the energy from this term
would be $E_{\beta}$
\begin{align}
E_{\beta} &  =\frac{4\pi}{V}\left[  \sum_{i=1}^{N-1}\sum_{j=i+1}^{N}q_{i}%
q_{j}\left(  \frac{1}{\beta^{2}}\right)  +\frac{1}{2}\sum_{i=1}^{N}q_{i}%
^{2}\left(  \frac{1}{\beta^{2}}\right)  \right]  \nonumber\\
&  =0
\end{align}
\qquad The result in Eq. (\ref{main})\ is the main result of this paper. The
form of $G$ as written in Eq. (\ref{main}) has excellent convergence for the
most part of the simulation cell. However, the convergence is bad for the case
when $\varepsilon=\rho_{0,0}/l_{1}\ll$ $1.$ The problem lies with the series
corresponding to $n=0$ and $p=0$ in Eq. (\ref{q}) for $Q$. For this case the
argument of the function $K_{0}$ becomes very small and it takes a lot of
summation terms over $m$ to achieve convergence. However, now there is now a
well defined way to fix this problem; we isolate the series corresponding to
$n=0$ and $p=0$, and rewrite it in terms of Polygamma and Zeta
functions\cite{tyagipre}. For this we first define%
\begin{align}
f\left(  x,\rho,l\right)  = &  \frac{1}{\sqrt{\rho^{2}+x^{2}}}+\sum
_{n=1}^{\infty}\left(  \frac{1}{\sqrt{\rho^{2}+\left(  nl+x\right)  ^{2}}%
}+\frac{1}{\sqrt{\rho^{2}+\left(  nl-x\right)  ^{2}}}-\frac{2}{nl}\right)
\nonumber\\
&  =\frac{4}{l}\sum_{m=1}^{\infty}\,K_{0}\left(  2\pi m\frac{\rho}{l}\right)
\cos\left(  2\pi m\frac{x}{l}\right)  -\frac{2}{l}\left\{  \gamma+\ln\left(
\frac{\rho}{2}\right)  \right\}  .\label{f}%
\end{align}
Using Eq. (\ref{main}) and (\ref{f}), we can write
\begin{align}
G\left(  \boldsymbol{r}\right)   &  =\frac{4}{l_{1}}\sum_{n,p}^{\prime}%
\sum_{m=1}^{\infty}\,K_{0}\left(  2\pi m\frac{\rho_{n,p}}{l_{1}}\right)
\cos\left(  2\pi m\frac{x_{n,p}}{l_{1}}\right)  \nonumber\\
&  -\frac{1}{l_{1}}\sum_{p}^{\prime}L\left[  \frac{x_{3}+l_{3}p}{l_{2}}%
\Omega,\frac{y_{p}}{l_{2}}\right]  +2\pi\left(  \frac{x_{3}^{2}}{l_{1}%
l_{2}l_{3}}-\frac{\left\vert x_{3}\right\vert }{l_{1}l_{2}}\right)
+\frac{2\gamma}{l_{1}}\nonumber\\
&  +f\left(  x_{0,0},\rho_{0,0},l_{1}\right)  +\frac{1}{l_{1}}\left\{
2\ln\left(  \frac{\rho_{0,0}}{2}\right)  -L\left[  \frac{x_{3}}{l_{2}}%
\Omega,\frac{y_{0}}{l_{2}}\right]  \right\}  ,\label{main1}%
\end{align}
where a prime over the summation sign indicates that the term corresponding to
$n=0$ and $p=0$ is not to be included in the summation. As written in Eq. (),
the series has two convergence problems when $\rho_{0,0}$ tends to zero.
Firstly, when $\rho_{0,0}=0$ the last term is not defined. And secondly,  as
mentioned earlier, the function $f\left(  x,\rho,l\right)  $ does not converge
fast for small $\rho.$ Both of  these problem may be fixed by taking the limit
$\rho_{0,0}\rightarrow0$ and combining the appropriate terms as follows. Using
Eq. (\ref{rhonp}) we note that $\rho_{0,0}\rightarrow0$ means $y_{0}%
\rightarrow0$ and $x_{3}\rightarrow0$. Also,%
\begin{equation}
\rho_{n,p}^{2}=\sin^{2}\alpha\left[  \left(  y_{p}+l_{2}n\right)  ^{2}+\left(
x_{3}+l_{3}p\right)  ^{2}\Omega^{2}\right]
\end{equation}
implies%
\begin{equation}
\rho_{0,0}^{2}=\sin^{2}\alpha\left[  y_{0}^{2}+x_{3}^{2}\Omega^{2}\right]  .
\end{equation}
So, the last term in Eq. (\ref{main1})\ may be written as%
\begin{align}
S  & =\frac{1}{l_{1}}\left\{  2\ln\left(  \frac{\rho_{0,0}}{2}\right)
-L\left[  \frac{x_{3}}{l_{2}}\Omega,\frac{y_{0}}{l_{2}}\right]  \right\}
\nonumber\\
& =\frac{1}{l_{1}}\left\{  \ln\left(  \frac{\sin^{2}\alpha\left[  y_{0}%
^{2}+x_{3}^{2}\Omega^{2}\right]  }{4l_{2}^{2}}l_{2}^{2}\right)  -L\left[
\frac{x_{3}}{l_{2}}\Omega,\frac{y_{0}}{l_{2}}\right]  \right\}  \nonumber\\
& =\frac{1}{l_{1}}\left\{  \ln\left(  \frac{\left[  y_{0}^{2}+x_{3}^{2}%
\Omega^{2}\right]  }{l_{2}^{2}}\right)  -L\left[  \frac{x_{3}}{l_{2}}%
\Omega,\frac{y_{0}}{l_{2}}\right]  -2\ln\left(  \frac{2}{l_{2}\left\vert
\sin\alpha\right\vert }\right)  \right\}  .
\end{align}
Now, through a simple Taylor expansion it could be shown\cite{sperb1} that for
small $y$ and $z$ we have%
\[
\ln\left(  y^{2}+z^{2}\right)  -L\left[  y,z\right]  =La\left[  y,z\right]
-2\ln\left(  2\pi\right)  ,
\]
where%
\begin{align}
La[y,z]  & =2\pi z+\frac{\pi^{2}}{3}\left(  y^{2}-z^{2}\right)  +\frac{\pi
^{4}}{90}\left(  y^{4}-6y^{2}z^{2}+z^{4}\right)  \nonumber\\
& +\frac{2\pi^{6}}{2835}\left(  y^{6}-15y^{4}z^{2}+15y^{2}z^{4}-z^{6}\right)
+\text{ higher order terms.}\label{sperb1}%
\end{align}
With the help of identity in Eq. (\ref{sperb1}) we can write $S$ for small
$\rho_{0,0}$ as follows:
\[
S=\frac{1}{l_{1}}La\left[  \frac{x_{3}}{l_{2}}\Omega,\frac{y_{0}}{l_{2}%
}\right]  -\frac{2}{l_{1}}\ln\left(  \frac{4\pi}{l_{2}\sin\alpha}\right)  .
\]
This fixes the first problem. To fix the second problem in order to achieve a
better convergence for small $\rho$, we re-express  $f\left(  x,\rho,l\right)
$ using the results of Ref. \cite{tyagipre} as follows:
\begin{align}
f\left(  x,\rho,l\right)   &  =\frac{1}{\left(  r_{1}^{2}+r_{2}^{2}+r_{3}%
^{2}+2r_{1}r_{2}\cos\alpha+2r_{2}r_{3}\cos\beta+2r_{3}r_{1}\cos\gamma\right)
^{1/2}}\nonumber\\
&  +\frac{1}{l}\sum_{n=1}^{N-1}\left(  \frac{1}{\sqrt{\rho^{2}+\left(
n+x\right)  ^{2}}}+\frac{1}{\sqrt{\rho^{2}+\left(  n-x\right)  ^{2}}}\right)
\nonumber\\
&  -\frac{2\gamma}{l}-\frac{\psi(N+x)+\psi(N-x)}{l}\nonumber\\
&  +\frac{1}{l}\sum_{m=1}^{\infty}\binom{-1/2}{m}\rho^{2m}\left[  \zeta\left(
2m+1,N+x\right)  +\zeta\left(  2m+1,N-x\right)  \right]  ,\label{deff}%
\end{align}
where $\psi$ and $\zeta$ stand for digamma and Hurwitz Zeta functions
respectively, and $N\geq1$ is chosen such that $N>\left(  \rho+x\right)  .$
However, for better convergence it is desirable that $N>(\rho+1)$. Thus, using
this alternate form for function $f$ from Eq. (\ref{deff}) we obtain a
different representation of $G$ which gives very fast convergence as $\rho$
tends to zero. The important fact here is that the Coulomb singularity toward
small $\left\vert \boldsymbol{r}\right\vert $ has been isolated. 

Also, it is now a simple matter to obtain the self-energy using the fact that%
\begin{align*}
\lim_{\rho_{0,0},x_{3}\rightarrow0}S  & =\lim_{y_{0}\rightarrow0,x_{3}%
\rightarrow0}\left(  \frac{1}{l_{1}}La\left[  \frac{x_{3}}{l_{2}}\Omega
,\frac{y_{0}}{l_{2}}\right]  -\frac{2}{l_{1}}\ln\left(  \frac{4\pi}{l_{2}%
\sin\alpha}\right)  \right)  \\
& =-\frac{2}{l_{1}}\ln\left(  \frac{4\pi}{l_{2}\sin\alpha}\right)  .
\end{align*}
It then follows
\begin{align}
G_{\text{self}} &  =\lim_{\rho,x\rightarrow0}\left(  G\left(  \boldsymbol{r}%
\right)  -\frac{1}{\sqrt{\rho^{2}+x^{2}}}\right)  \nonumber\\
&  =\frac{4}{l_{1}}\sum_{n,p}^{\prime}\sum_{m=1}^{\infty}\,K_{0}\left(  2\pi
m\frac{\rho_{n,p}^{0}}{l_{1}}\right)  \times\cos\left(  2\pi m\frac
{x_{n,p}^{0}}{l_{1}}\right)  +\frac{2\gamma}{l_{1}}\nonumber\\
&  -\frac{1}{l_{1}}\sum_{p}^{\prime}L\left[  \frac{l_{3}p}{l_{2}}\Omega
,\frac{\zeta l_{3}p}{l_{2}}\right]  -\frac{2}{l_{1}}\ln\left(  \frac{4\pi
}{l_{2}\sin\alpha}\right)  ,\label{self}%
\end{align}
where%
\begin{align*}
\rho_{n,p}^{0}  & =\sqrt{\sin^{2}\alpha\left[  \left(  l_{3}p\zeta
+l_{2}n\right)  ^{2}+\left(  l_{3}p\right)  ^{2}\Omega^{2}\right]  }.\\
\end{align*}
Expressions in Eq. (\ref{main}) and (\ref{self}) are the generalization of
Sperb's and Lekner's work from an orthorhombic cell to a a triclinic cell. In
the special case when a orthorhombic cell is considered, Eq. (\ref{main})
reduces to
\begin{align}
G_{\text{ortho}} &  =\frac{4}{l_{1}}\sum_{n,p}^{\prime}\sum_{m=1}^{\infty
}\,K_{0}\left(  2\pi m\frac{\sqrt{\left(  x_{2}+l_{2}n\right)  ^{2}+\left(
x_{3}+l_{3}p\right)  ^{2}}}{l_{1}}\right)  \cos\left(  2\pi m\frac{x_{1}%
}{l_{1}}\right)  \nonumber\\
&  -\frac{1}{l_{1}}\sum_{p=-\infty}^{\infty}\ln\left[  1-2\exp\left(
-\frac{2\pi}{l_{2}}\left\vert x_{3}+l_{3}p\right\vert \right)  \right.
\nonumber\\
&  \left.  \times\cos\left(  2\pi\frac{x_{2}}{l_{2}}\right)  +\exp\left(
-\frac{4\pi}{l_{2}}\left\vert x_{3}+l_{3}p\right\vert \right)  \right]
\nonumber\\
&  +\frac{2}{l_{1}}\ln\left(  \sqrt{x_{2}^{2}+x_{3}^{2}}\right)  +f\left(
x_{1},\sqrt{x_{2}^{2}+x_{3}^{2}},l_{1}\right)  -2\pi\frac{\left\vert
x_{3}\right\vert }{l_{1}l_{2}}+2\pi\frac{x_{3}^{2}}{l_{1}l_{2}l_{3}}.
\end{align}
This result is in agreement with the ones derived by Sperb and Lekner.
Finally, to summarize our results, the total energy of $N$ charges is given by
Eq. (\ref{tot}). The function $G$ in Eq. (\ref{tot}) may be obtained by using
\begin{align}
G\left(  \boldsymbol{r}\right)   &  =\frac{4}{l_{1}}\sum_{n,p}\sum
_{m=1}^{\infty}\,K_{0}\left(  2\pi m\frac{\rho_{n,p}}{l_{1}}\right)
\cos\left(  2\pi m\frac{x_{n,p}}{l_{1}}\right)  -\frac{1}{l_{1}}%
\sum_{p=-\infty}^{\infty}L\left[  \frac{x_{3}+l_{3}p}{l_{2}}\Omega,\frac
{y_{p}}{l_{2}}\right]  \nonumber\\
&  +\frac{\Omega l_{3}\sin\alpha}{l_{1}l_{2}}\frac{\pi}{3}\left[
1-6\frac{\left\vert x_{3}\right\vert }{l_{3}}+\left(  \frac{x_{3}}{l_{3}%
}\right)  ^{2}\right]  ,
\end{align}
where $\rho_{n,p}$, $x_{n,p}$, $y_{p}$ are defined in Eqs. (\ref{defrho}),
(\ref{defx}), and (\ref{defy}). The $\Omega$ is defined by Eq. (\ref{defomega}%
) and the function $L$ is defined in Eq. (\ref{defL}). For the case when
$\varepsilon=\rho_{0,0}/l_{1}\ll$ $1$, one should use the following
expression
\begin{align}
G\left(  \boldsymbol{r}\right)   &  =\frac{4}{l_{1}}\sum_{n,p}^{\prime}%
\sum_{m=1}^{\infty}\,K_{0}\left(  2\pi m\frac{\rho_{n,p}}{l_{1}}\right)
\cos\left(  2\pi m\frac{x_{n,p}}{l_{1}}\right)  \nonumber\\
&  -\frac{1}{l_{1}}\sum_{p}^{\prime}L\left[  \frac{x_{3}+l_{3}p}{l_{2}}%
\Omega,\frac{y_{p}}{l_{2}}\right]  +2\pi\left(  \frac{x_{3}^{2}}{l_{1}%
l_{2}l_{3}}-\frac{\left\vert x_{3}\right\vert }{l_{1}l_{2}}\right)
\nonumber\\
&  +f\left(  x_{0,0},\rho_{0,0},l_{1}\right)  +\frac{2\gamma}{l_{1}}+\frac
{1}{l_{1}}La\left[  \frac{x_{3}}{l_{2}}\Omega,\frac{y_{0}}{l_{2}}\right]
-\frac{2}{l_{1}}\ln\left(  \frac{4\pi}{l_{2}\sin\alpha}\right)  ,
\end{align}
where and $f$ is is defined in Eq. (\ref{deff}) and $La$ is defined in Eq
.(\ref{sperb1}) . The expression for force may be obtained by using
$\boldsymbol{F}\left(  \boldsymbol{r}\right)  =-\boldsymbol{\nabla}G\left(
\boldsymbol{r}\right)  $. Finally, the self-energy required in Eq. (\ref{tot})
is obtained by using%
\begin{align}
G_{\text{self}} &  =\frac{4}{l_{1}}\sum_{n,p}^{\prime}\sum_{m=1}^{\infty
}\,K_{0}\left(  2\pi m\frac{\rho_{n,p}^{0}}{l_{1}}\right)  \times\cos\left(
2\pi m\frac{x_{n,p}^{0}}{l_{1}}\right)  \nonumber\\
&  -\frac{1}{l_{1}}\sum_{p}^{\prime}L\left[  \frac{l_{3}p}{l_{2}}\Omega
,\frac{\zeta l_{3}p}{l_{2}}\right]  -\frac{2}{l_{1}}\ln\left(  \frac{4\pi
}{l_{2}\left\vert \sin\alpha\right\vert }\right)  .
\end{align}

\section{Results and Conclusions}

We have obtained complete expressions for the Coulomb potential in 3D,
including self-energies. The results were derived for the most general case of
a triclinic cell in 3D. The formulas derived here are in the same form as
derived earlier by Lekner\cite{lekner} and Sperb\cite{sperb1} for the case of
an orthorhombic cell. With our results, it should now be possible to extend
any written code for Lekner or Sperb summation to a triclinic cell with
minimal effort. The results obtained in this paper reduces to the results of a
recent paper\cite{tyagipre} when all angles pertaining to the unit cell are
set to $\pi/2$. The results also agree with the special cases considered by
Lekner\cite{lekner} and Sperb\cite{sperb1,mmm}.

\appendix

\section{Representation for $G(\boldsymbol{r};\beta)$}

The solution of
\begin{equation}
\left(  \nabla^{2}-\beta^{2}\right)  S_{d}(\boldsymbol{r};\beta)=-C_{d}%
\delta_{d}(\boldsymbol{r})\label{one}%
\end{equation}
in $d$-dimensional space is given by%
\begin{equation}
S_{d}(\boldsymbol{r};\beta)=\frac{C_{d}}{\left(  2\pi\right)  ^{\nu+1/2}%
}\left[  \beta^{\nu}\frac{K_{\nu}\left(  \beta r\right)  }{r^{\nu}}\right]
,\label{a14}%
\end{equation}
where $C_{d}$ is defined as%
\[
C_{d}=\left\{
\begin{array}
[c]{c}%
2\ \ \ \ \ \ \ \ \ \ \ \ \ \ \ \ \text{~\ }d=1\\
2\pi\text{\ \ \ \ \ \ \ \ \ \ \ \ \ \ \ \ \ }d=2,\\
4\pi^{\nu+1}/\Gamma\left(  \nu\right)  \text{ \ \ \ }d>2.
\end{array}
\right.
\]
Here, $\Gamma(\nu)$ stands for the Gamma function, and $\nu=\left(
d-2\right)  /2$. For $d=3$ we can write the solution in the Fourier space
using Eq. (\ref{one}):%
\begin{equation}
S_{3}(\boldsymbol{r};\beta)=\frac{C_{3}}{\left(  2\pi\right)  ^{2}}\int
dk_{1}d\boldsymbol{k}_{\rho}\frac{e^{i2\pi(k_{1}x+\boldsymbol{k}_{\rho
}.\boldsymbol{\rho})}}{k_{1}^{2}+\boldsymbol{k}_{\rho}^{2}+\beta^{2}/4\pi^{2}%
},
\end{equation}
where we have written the vector $\boldsymbol{r}$ in cylindrical coordinates
as $\left(  x,\boldsymbol{\rho}\right)  .$ Now, using Eq. (\ref{g2}) we can
write%
\begin{align}
G\left(  \boldsymbol{r};\beta\right)   &  =\frac{C_{3}}{\left(  2\pi\right)
^{3/2}}\beta^{1/2}\sum_{m,n,p}\frac{K_{1/2}\left(  \beta r_{m,n,p}\right)
}{r_{m,n,p}^{1/2}}\nonumber\\
&  =\frac{C_{3}}{\left(  2\pi\right)  ^{2}}\sum_{n,p}\sum_{m}\int
dk_{1}d\boldsymbol{k}_{\rho}\frac{e^{i2\pi\left[  k_{1}\left(  x_{n,p}%
+l_{1}m\right)  +\boldsymbol{k}_{\rho}.\boldsymbol{\rho})\right]  }}{k_{1}%
^{2}+\boldsymbol{k}_{\rho}^{2}+\beta^{2}/4\pi^{2}}.
\end{align}
Now, using the fact that
\begin{equation}
\sum_{m}e^{i2\pi k_{1}l_{1}m}=\frac{1}{l_{1}}\sum_{m}\delta(k_{1}-m),
\end{equation}
we can replace the integral over $k_{1}$ with a summation:%
\begin{align}
G\left(  \boldsymbol{r};\beta\right)   &  =\frac{C_{3}}{\left(  2\pi\right)
^{2}}\frac{1}{l_{1}}\sum_{n,p}\sum_{m}\int d\boldsymbol{k}_{\rho}%
\frac{e^{i2\pi(mx_{n,p}+\boldsymbol{k}_{\rho}.\boldsymbol{\rho}_{n,p})}}%
{m^{2}+\boldsymbol{k}_{\rho}^{2}+\beta^{2}/4\pi^{2}}\nonumber\\
&  =\frac{C_{3}}{C_{2}}\frac{1}{l_{1}}\sum_{n,p}\sum_{m}\exp\left(  i2\pi
m\frac{x_{np}}{l_{1}}\right)  K_{0}\left(  \sqrt{m^{2}+\beta^{2}}\rho
_{n,p}\right)  ,
\end{align}
where
\[
\rho_{n,p}=\left\vert \boldsymbol{\rho}_{n,p}\right\vert
\]

\bigskip

\begin{acknowledgments}
Support from the Volkswagen foundation, grant number I/80 433, is gratefully acknowledged.
\end{acknowledgments}

\end{document}